\documentclass[doublecol]{epl2} 
\usepackage{graphicx}
\usepackage{color}
\usepackage{amsmath}
\usepackage{amssymb}
\usepackage{bm}

\newcommand{\Tr}{\mathop{\text{Tr}}\nolimits}
\newcommand{\ket}[1]{|{#1}\rangle}
\newcommand{\bra}[1]{\langle{#1}|}
\newcommand{\bras}[2]{{}_{#2}\hspace*{-0.2mm}\langle{#1}|}

\newcommand{\pv}{\mathop{\text{P}}\nolimits}

\renewcommand{\Im}{\mathop{\text{Im}}\nolimits}

\definecolor{dgreen}{rgb}{0,0.5,0}

\definecolor{delete}{cmyk}{0.5,0,0,0}

\newcommand{\DEL}[1]{}

\title{Typical pure nonequilibrium steady states}
\shorttitle{Typical pure nonequilibrium stationary states} 

\author{Takaaki Monnai\inst{1}\thanks{E-mail: \email{monnai@suou.waseda.jp}} \and Kazuya Yuasa\inst{2}\thanks{E-mail: \email{yuasa@waseda.jp}}}
\shortauthor{Takaaki Monnai and Kazuya Yuasa}

\institute{                    
  \inst{1} Waseda Institute for Advanced Study, Waseda University, Tokyo 169-8050, Japan\\
  \inst{2} Department of Physics, Waseda University, 
Tokyo 169-8555, Japan
}

\abstract{
We show that typicality holds for a class of nonequilibrium systems, i.e., nonequilibrium steady states (NESSs): almost all the pure states properly sampled from a certain Hilbert space well represent a NESS and characterize its intrinsic thermal nature.
We clarify the relevant Hilbert space from which the pure states are to be sampled, and  construct practically all the typical pure NESSs.
The scattering approach leads us to the natural extension of the typicality for equilibrium systems.
Each pure NESS correctly yields the expectation values of observables given by the standard ensemble approach.
It means that we can calculate the expectation values in a NESS with only a single pure NESS\@.
We provide an explicit construction of the typical pure NESS for a model with two reservoirs, and see that it correctly reproduces the Landauer-type formula for the current flowing steadily between the reservoirs.
}

\begin{document}
\maketitle

\section{Introduction}
Recently, the \textit{typicality} of pure states of large quantum systems has been attracting considerable attentions for the foundations of statistical mechanics. 
It has been shown that the partially reduced states of the overwhelming majority of the pure states in an energy shell of a large quantum system look similar to each other, and well approximate the canonical state (``canonical typicality''), for which the importance of a large entanglement in each typical pure state of the whole system is stressed \cite{Lebowitz1,Popescu1} (see also \cite{ref:TasakiCanonicalTypicality,ref:GemmerEPJB}).
It has been also pointed out that a similar typicality already holds for the whole system: almost all the pure states in the energy shell of the whole system give the expectation values of an observable very close to that evaluated in the microcanonical state \cite{Sugita1b,Sugita1,Reimann1}. 
Such intrinsic thermal nature of typical pure states enables us to analyze mechanical and thermodynamical quantities \cite{Sugiura1,Sugiura2} and their fluctuations, especially the probability distributions \cite{Monnai1}, by only a single pure state. 
The equilibration/thermalization \cite{Rigol1,Reimann2,Popescu2,Lebowitz2,Short1,Reimann3}, the temporal fluctuation around equilibrium \cite{Popescu3}, and the relaxation time \cite{Goldstein1,Monnai2} have also been discussed for large quantum systems on the basis of the typicality arguments, which would be relevant to the experimentally observed emergence of thermal correlations in an isolated cold atomic gas \cite{ref:Schmidmayer-AtomChip-NaturePhys9}.
Instead of the statistical ansatz such as \textit{a priori} equal probabilities of energy eigenstates and the erogodicity hypothesis, the quantum spectral fluctuation provides alternative probabilistic properties necessary for the statistical mechanics.

These extraordinary achievements for typical pure states being at hand, it is natural to expect that the typicality holds also for nonequilibrium systems.
As a first step, we are going to focus on nonequilibrium steady states (NESSs) \cite{NESS-Ruelle,Tasaki3,Tasaki2,ref:NESS-AschbacherPillet-JSP,NESS-AschbacherJaksicPautratPillet,NESS-Tasaki,Gaspard1,Saito1}.
Needless to say, the NESSs are of both practical and fundamental importance: they are of course relevant to the DC conductivity \cite{Mahan1}, and include interesting topics such as the long-range correlations in the NESSs \cite{Antal1}.
On the other hand, the universal properties of the NESSs are less understood \cite{Antal1,Tasaki2} compared with those of equilibrium systems.
In addition, it is not clear whether the typicality holds for NESSs and how to construct typical pure states representing NESSs if any.

In this paper, we give an answer to this last question.
We provide a construction of practically all \textit{typical pure NESSs}.
Our idea is to reduce the problem for the nonequilibrium situation to that for the equilibrium through the Lippmann-Schwinger scattering theory \cite{Thirring1,ScatteringTaylor}: we ``scatter'' a typical pure state representing an equilibrium state to generate a typical pure NESS\@.
We show that it actually gives the same expectation value as the standard ensemble approach.
For concreteness, we investigate the simplest model for a quantum transport, and give an explicit construction of a typical pure NESS\@.

\section{Typicality for equilibrium systems}
Let us first recapitulate an aspect of the typicality for equilibrium systems, relevant to the following discussion.
We consider a large quantum system, and an energy shell $\mathcal{H}_E$ spanned by the energy eigenstates $\{\ket{E_i}\}$ belonging to the energies between $E$ and $E+\Delta E$.
The energy width $\Delta E$ is supposed to be small, but the system is so large that the energy shell $\mathcal{H}_E$ contains many energy eigenstates $\{\ket{E_i}\}$, i.e., $d=\dim\mathcal{H}_E$ is large.
We then pick a pure state \cite{Lebowitz1,Popescu1,Sugita1b,Sugita1,Reimann1,Sugiura1}
\begin{equation}
\ket{\psi}=\sum_{i=1}^dc_i\ket{E_i}
\label{eqn:TypicalPureStateEq}
\end{equation}
randomly from the energy shell $\mathcal{H}_E$, i.e., the complex coefficients $\{c_i\}$ are chosen from the uniform distribution on the surface of the $2d$-dimensional unit sphere $\sum_{i=1}^d|c_i|^2=1$, according to the Haar measure.
Such a \textit{single} pure state $\ket{\psi}$ \textit{typically} well represents the microcanonical ensemble on the energy shell $\mathcal{H}_E$:
almost all the pure states $\ket{\psi}$ on the energy shell $\mathcal{H}_E$ give the expectation values $\bra{\psi}\hat{A}\ket{\psi}$ of an observable $\hat{A}$ close to the expectation value $\langle\hat{A}\rangle_\text{mc}=\Tr\{\hat{\rho}_\text{mc}\hat{A}\}$ in the microcanonical ensemble $\hat{\rho}_\text{mc}=\frac{1}{d}\sum_{i=1}^d\ket{E_i}\bra{E_i}$ \cite{Sugita1b,Sugita1,Reimann1,Sugiura1},
\begin{equation}
\bra{\psi}\hat{A}\ket{\psi}\sim\langle\hat{A}\rangle_\text{mc}.
\end{equation}
More rigorously, the probability of $\bra{\psi}\hat{A}\ket{\psi}$ deviating from $\langle\hat{A}\rangle_\text{mc}$ is bounded, for any positive $K$, by
\begin{equation}
P\!\left(|\langle\psi|\hat{A}|\psi\rangle-\langle\hat{A}\rangle_\text{mc}|^2> K\frac{(\Delta\hat{A})_\text{mc}^2}{d+1}\right)<\frac{1}{K},
\label{inequality1}
\end{equation}
where $(\Delta\hat{A})_\text{mc}^2=\langle\hat{A}^2\rangle_\text{mc}-\langle\hat{A}\rangle_\text{mc}^2$.  
Roughly, it means that the error is typically $\mathcal{O}((\Delta\hat{A})_\text{mc}/\sqrt{d})$ \cite{Monnai1}.
This is valid even for an unbounded operator $\hat{A}$ as long as the microcanonical variance $(\Delta\hat{A})_\text{mc}^2$ is under control.
Each typical pure state $\ket{\psi}$ on the energy shell $\mathcal{H}_E$ well describes the thermal equilibrium state of the system.

Note that we do not put the system in contact with an ``environment.'' Such an external agent is not necessary for the typicality and for the equivalence with the microcanonical state \cite{Sugita1b,Sugita1,Reimann1,Sugiura1}. If we divide the system into two parts and take the partial trace over one of them, all the reduced states of the typical pure states in the energy shell $\mathcal{H}_E$ of the total system are very close to a canonical state. It is called ``canonical typicality'' \cite{Lebowitz1,Popescu1} (see also \cite{ref:TasakiCanonicalTypicality,ref:GemmerEPJB}). We are not going to discuss this in this paper: we will not have an environment and will not take a partial trace.

The objective of the present paper is to extend this typicality argument to a nonequilibrium situation, i.e., to NESS\@.
At first glance, it is not clear from which subspace we should sample pure states.
Moreover, there is no clear clue whether the typicality holds also for nonequilibrium systems. 
For NESSs, nonetheless, we will clarify that the typicality actually holds: we will identify the relevant subspace and construct practically all \textit{pure} NESSs.

\section{Nonequilibrium steady state (NESS)} 
\begin{figure}
\centerline{
\includegraphics[width=0.34\textwidth]{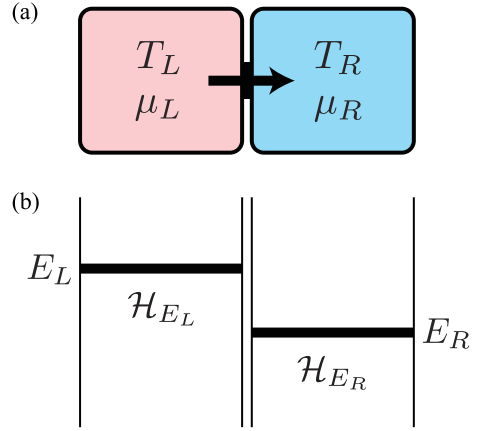}
}
\caption{(a) A reservoir at temperature $T_L$ with chemical potential $\mu_L$ is connected via a weak local link with another reservoir at temperature $T_R$ with chemical potential $\mu_R$, and a current flows steadily between them.
(b) The initial pure state is randomly sampled from the subspace $\mathcal{H}_{E_L,E_R}=\mathcal{H}_{E_L}\otimes\mathcal{H}_{E_R}$, where $\mathcal{H}_{E_L}$ and $\mathcal{H}_{E_R}$ are the energy shells around $E_L$ and $E_R$ in the Hilbert spaces of reservoirs $L$ and $R$, respectively.
}
\label{fig:HirbertNESS}
\end{figure}

The simplest setup for discussing the NESS is shown in Fig.\ \ref{fig:HirbertNESS}(a) \cite{Tasaki2,Tasaki3,NESS-Ruelle,ref:NESS-AschbacherPillet-JSP,NESS-AschbacherJaksicPautratPillet,NESS-Tasaki,Gaspard1,Saito1,Mahan1}.\footnote{We can easily generalize the following arguments to the systems with multiple reservoirs, but we focus on the two-reservoir case for simplicity.}
We have two reservoirs $L$ and $R$, one at temperature $T_L$ with chemical potential $\mu_L$ and the other at $T_R$ with $\mu_R$.
Connecting them via a weak link, a current starts to flow between them.
This can be described in the following natural way \cite{Tasaki2,Tasaki3,NESS-Ruelle,ref:NESS-AschbacherPillet-JSP,NESS-AschbacherJaksicPautratPillet,NESS-Tasaki,Gaspard1,Saito1,Mahan1}: we let the whole system evolve from the initial state $\hat{\rho}_L\otimes\hat{\rho}_R$ by a Hamiltonian 
\begin{equation}
\hat{H}=\hat{H}_0+\hat{V},\qquad
\hat{H}_0=\hat{H}_L+\hat{H}_R,
\label{eqn:Hamiltonian}
\end{equation}
where $\hat{H}_L$ and $\hat{H}_R$ are the Hamiltonians of the reservoirs $L$ and $R$, respectively, with a weak local interaction $\hat{V}$ which transfers particles between the two reservoirs. $\hat{\rho}_L$ and $\hat{\rho}_R$ represent the thermal equilibrium states of the reservoirs $L$ and $R$, respectively, with the relevant temperatures and chemical potentials. Usually, the grand canonical ensembles $\hat{\rho}_\text{gc}^{(L(R))}=e^{-\beta_{L(R)}(\hat{H}_{L(R)}-\mu_{L(R)}\hat{N}_{L(R)})}/\Xi_{L(R)}$ are taken, with $\beta_{L(R)}$ being the inverse temperature, $\hat{N}_{L(R)}$ the number of particles in reservoir $L(R)$, and $\Xi_{L(R)}$ the grand canonical partition function.

If the two systems $L$ and $R$ are large but of finite size, the whole system eventually reaches an equilibrium state with the two systems at the same temperature. Here, instead, we are interested in infinitely large systems as $L$ and $R$, and the thermodynamical limit will be taken. Then, in the long-time limit, the system reaches a NESS, in which a current flows steadily. We are going to discuss such NESSs \cite{Tasaki2,Tasaki3,NESS-Ruelle,ref:NESS-AschbacherPillet-JSP,NESS-AschbacherJaksicPautratPillet,NESS-Tasaki,Gaspard1,Saito1,Mahan1}.

Mathematically the state $e^{-i\hat{H}t}(\hat{\rho}_L\otimes\hat{\rho}_R)e^{i\hat{H}t}$ does not converge in general (more rigorously, the expectation values of observables in this state would not converge) in the long-time limit $t\to\infty$.
The mathematically well-defined construction of NESS instead is given by \cite{Tasaki2,NESS-Ruelle,ref:NESS-AschbacherPillet-JSP,NESS-AschbacherJaksicPautratPillet,NESS-Tasaki}
\begin{equation}
\hat{\rho}_\text{NESS}
=\hat{\Omega}(\hat{\rho}_L\otimes\hat{\rho}_R)\hat{\Omega}^\dag
\label{eqn:RhoNESS}
\end{equation}
with the M\o ller wave operator \cite{Thirring1,ScatteringTaylor}
\begin{equation}
\hat{\Omega}=\lim_{t\rightarrow\infty}e^{-i\hat{H} t}e^{i\hat{H}_0 t}.
\label{eqn:Omega}
\end{equation}
The very rough idea is to remove the free oscillations by $e^{i\hat{H}_0t}$ to ensure the convergence in the long-time limit.
In the context of a scattering problem, the wave operator $\hat{\Omega}$ transforms incident plane waves to the scattering states satisfying the Lippmann-Schwinger equation under the outgoing boundary condition.
In the present context, the free state $\hat{\rho}_L\otimes\hat{\rho}_R$ of $\hat{H}_0$ as the incident state is transformed into a scattering state by $\hat{\Omega}$, which is the NESS\@.
We are interested, in particular, in the current between the two reservoirs.
We compute the expectation value of the current operator (see the example below), which approaches a well-defined stationary value in the long-time limit: the current flows steadily, and this is a NESS\@.

There might exist some pathological system for which the long-time limit in (\ref{eqn:Omega}) does not exist. In the following, we exclude such an exceptional case and assume the existence of a well-defined NESS\@.

\section{Construction of typical pure NESSs}
Having seen the above approach to the NESS, we are led to the following natural way for constructing \textit{typical pure NESSs}.
The basic idea is to sample a typical pure state representing the thermal \textit{equilibrium} states $\hat{\rho}_L$ and $\hat{\rho}_R$ of the reservoirs, and ``scatter'' it by the wave operator $\hat{\Omega}$.

More specifically, we take a subspace $\mathcal{H}_{E_L,E_R}=\mathcal{H}_{E_L}\otimes\mathcal{H}_{E_R}$ in the Hilbert space of the total system, where $\mathcal{H}_{E_{L(R)}}$ is the energy shell around the energy $E_{L(R)}$ of reservoir $L(R)$ with a small energy width $\Delta E$, spanned by the energy eigenstates $\{\ket{E_i^{(L(R))}}\}$ in the energy range [see Fig.\ \ref{fig:HirbertNESS}(b)].
Its dimension is $D=\dim\mathcal{H}_{E_L,E_R}=d_Ld_R$ with $d_{L(R)}=\dim\mathcal{H}_{E_{L(R)}}$.\footnote{\label{Note3}For the discussion of the typicality, we assume the volumes of the reservoirs to be large but finite, so that $d_{L(R)}$ are finite. The thermodynamical limit is to be taken later, after the expectation values of observables are computed. Note that the long-time limit in (\ref{eqn:Omega}) requires a continuous spectrum. This limit is understood to be taken after the large-volume limit. Before it, the time $t$ is kept sufficiently large but finite. Note also that the current remains finite in the thermodynamical limit since it flows by a local interaction $\hat{V}$.}
We then pick a pure state $\ket{\phi}$ randomly from the subspace $\mathcal{H}_{E_L,E_R}$ with respect to the Haar measure, and define
\begin{equation}
|\phi\rangle_\infty=\hat{\Omega}|\phi\rangle.
\label{stationary1}
\end{equation}
Our main result is that the pure state $|\phi\rangle_\infty$ constructed in this way is a \textit{typical pure  NESS} in the well-defined thermodynamical limit.
That is, almost all the pure states $\ket{\phi}_\infty$ sampled in this way give the expectation values $\bras{\phi}{\infty}\hat{A}\ket{\phi}_\infty$ close to the expectation value $\langle\hat{A}\rangle_\text{NESS}=\Tr\{\hat{\rho}_\text{NESS}\hat{A}\}$ evaluated with $\hat{\rho}_\text{NESS}$ given in (\ref{eqn:RhoNESS}),
\begin{equation}
\bras{\phi}{\infty}\hat{A}\ket{\phi}_\infty\sim\langle\hat{A}\rangle_\text{NESS}.
\label{eqn:PureNESSAve}
\end{equation}
Note that $|\phi\rangle_\infty$ is invariant under the time evolution up to a global phase, since 
$e^{-i\hat{H}\tau}\hat{\Omega}|\phi\rangle=\hat{\Omega} e^{-i\hat{H}_0\tau}|\phi\rangle$ holds for an arbitrary $\tau$ \cite{ScatteringTaylor} and $|\phi\rangle$ is approximately an eigenstate of $\hat{H}_0$ due to the assumption of the small energy width $\Delta E$.
Thus, the expectation values in $\ket{\phi}_\infty$ are stationary.

Before proving the typicality of $\ket{\phi}_\infty$, it is important to point out the following fact.
One distinguished characteristics of NESS, in contrast to equilibrium cases, is that it is endowed with multiple different temperatures and chemical potentials, or in other words, it is characterized by multiple different energy scales.
We basically intend to take a typical pure state $\ket{\phi}$ representing the product of the two thermal equilibrium states $\hat{\rho}_L\otimes\hat{\rho}_R$ as the initial state, but our sampling is not performed separately for $\mathcal{H}_{E_L}$ and $\mathcal{H}_{E_R}$.
Each initial state $\ket{\phi}$ is sampled from $\mathcal{H}_{E_L,E_R}$, which in general is not of the product structure, but rather \textit{entangled},
\begin{equation}
\ket{\phi}=\sum_{i=1}^{d_L}\sum_{j=1}^{d_R}c_{ij}\ket{E_i^{(L)}}\otimes\ket{E_j^{(R)}}.
\label{eqn:Entanglement}
\end{equation}
It still typically represents the product state $\hat{\rho}_L\otimes\hat{\rho}_R$, as we will see below.
This is an important point.
We are exploring the whole of the subspace $\mathcal{H}_{E_L,E_R}$ characterized by the two intrinsic energy scales $E_L$ and $E_R$.
Then, through (\ref{stationary1}), which establishes an isometric construction of the pure NESSs $\ket{\phi}_\infty$ expressed as scattering states, practically all pure NESSs are at our hand.\footnote{The wave operator $\hat{\Omega}$ is not unitary in the presence of nondecaying bound states $\{{\ket{\phi_{B,i}}}\}$ \cite{Thirring1,ScatteringTaylor}. Indeed, $\hat{\Omega}^\dag\hat{\Omega}=1$ but $\hat{\Omega}\hat{\Omega}^\dag=1-\sum_i{|\phi_{B,i}\rangle}{\langle\phi_{B,i}}|$. Equation (\ref{stationary1}) further shows that ${|\phi\rangle}_\infty=(1-\sum_i{|\phi_{B,i}\rangle}{\langle\phi_{B,i}|}){|\phi\rangle}_\infty$. This means that the pure NESS ${|\phi\rangle}_\infty$ constructed by (\ref{stationary1}) is necessarily a scattering state. For the discussion of NESSs, it is reasonable to exclude the superpositions of energetically separated scattering and bound states.}

\section{Typicality of pure NESSs}
The proof of the typicality of the pure NESSs $\ket{\phi}_\infty$ consists of two steps.
The first step is to show
\begin{equation}
\mathbb{E}[\bra{\phi}\hat{\Omega}^\dag\hat{A}\hat{\Omega}\ket{\phi}]
=\Tr\{(\hat{\rho}_\text{mc}^{(L)}\otimes\hat{\rho}_\text{mc}^{(R)})\hat{\Omega}^\dag\hat{A}\hat{\Omega}\},
\label{eqn:AveMC}
\end{equation}
where $\mathbb{E}[{}\cdots{}]$ represents the ensemble average over the uniformly sampled $\ket{\phi}$ with respect to the Haar measure in $\mathcal{H}_{E_L,E_R}$, and $\hat{\rho}_\text{mc}^{(L(R))}=\frac{1}{d_{L(R)}}\sum_{i=1}^{d_{L(R)}}\ket{E_i^{(L(R))}}\bra{E_i^{(L(R))}}$ is the microcanonical ensemble for the thermal equilibrium state of reservoir $L(R)$.
We abbreviate it to $\Tr\{(\hat{\rho}_\text{mc}^{(L)}\otimes\hat{\rho}_\text{mc}^{(R)})\hat{\Omega}^\dag\hat{A}\hat{\Omega}\}\equiv\langle\hat{\Omega}^\dag\hat{A}\hat{\Omega}\rangle_\text{mc}$.
Due to the equivalence among the microcanonical, canonical, and grand canonical ensembles in the thermodynamical limit,\footnote{This is valid only in the absence of condensation \cite{Holthaus1998198,PhysRevLett.112.030401}. We exclude cases with symmetry breaking, where the grand canonical catastrophe occurs. Such cases will be studied elsewhere.}
the above microcanonical average $\langle\hat{\Omega}^\dag\hat{A}\hat{\Omega}\rangle_\text{mc}$ well coincides with the grand canonical average $\Tr\{(\hat{\rho}_\text{gc}^{(L)}\otimes\hat{\rho}_\text{gc}^{(R)})\hat{\Omega}^\dag\hat{A}\hat{\Omega}\}$, which is ${\langle\hat{A}\rangle}_\text{NESS}$ in (\ref{eqn:PureNESSAve}) usually computed in the literature.
We then see that the variance is bounded by
\begin{equation}
\mathbb{V}[\langle\phi|\hat{\Omega}^\dag\hat{A}\hat{\Omega}|\phi\rangle]
\le\frac{
(\Delta(\hat{\Omega}^\dag\hat{A}\hat{\Omega}))^2_\text{mc}
}{D+1}
\label{secondorder1}
\end{equation}
with $\mathbb{V}[A]=\mathbb{E}[A^2]-\mathbb{E}[A]^2$, $(\Delta\hat{A})_\text{mc}^2=\langle\hat{A}^2\rangle_\text{mc}-\langle\hat{A}\rangle_\text{mc}^2$, and that it
shrinks as $D\rightarrow\infty$, ensuring the typicality (\ref{eqn:PureNESSAve}).

The crucial observation is that, thanks to the introduction of the wave operator $\hat{\Omega}$, the expectation value of $\hat{A}$ in the NESS is translated into the expectation value of $\hat{\Omega}^\dag\hat{A}\hat{\Omega}$ in the double thermal equilibrium state.
Therefore, we can easily extend the standard arguments for equilibrium systems \cite{Sugita1b,Sugita1,Reimann1} to the present discussion for NESSs.
One subtle nontrivial point is the entanglement in (\ref{eqn:Entanglement}) to represent the product of the two thermal equilibrium states $\hat{\rho}_L\otimes\hat{\rho}_R$.
It is however not a difficult problem.

Here comes the proof of the typicality of the pure NESSs $\ket{\phi}_\infty$.
The key formulas are available by generalizing those for equilibrium cases with the single-index coefficients $\{c_i\}$ in (\ref{eqn:TypicalPureStateEq}) to the multi-index coefficients $\{c_{ij}\}$ in (\ref{eqn:Entanglement}).
The Haar measure is proportional to $\delta(\sum_{i=1}^{d_L}\sum_{j=1}^{d_R}|c_{ij}|^2-1)\prod_{i=1}^{d_L}\prod_{j=1}^{d_R}d^2c_{ij}$, which yields \cite{Sugita1b,Sugita1,Ullah196465,ref:GemmerTextbook}
\begin{equation}
\mathbb{E}[|c_{ij}|^2]
=
\frac{1}{D},\quad
\mathbb{E}[|c_{ij}|^2|c_{i'j'}|^2]
=
\frac{1+\delta_{ii'}\delta_{jj'}}{D(D+1)},
\end{equation}
while the others up to the fourth moments of $c_{ij}$ vanish.
Using these formulas, we get for any operator $\hat{A}$
\begin{equation}
\mathbb{E}[\bra{\phi}\hat{A}\ket{\phi}]
=\frac{1}{D}
\sum_{i=1}^{d_L}\sum_{j=1}^{d_R}A_{ij,ij}
=\langle\hat{A}\rangle_\text{mc}
\label{eqn:MicroPhi}
\end{equation}
and
\begin{align}
&\mathbb{V}[\langle\phi|\hat{A}|\phi\rangle]
=\frac{1}{D(D+1)}
\sum_{i,i'=1}^{d_L}\sum_{j,j'=1}^{d_R}
|A_{ij,i'j'}|^2
\nonumber\\
&\qquad\quad
{}-\frac{1}{D^2(D+1)}\left(\sum_{i=1}^{d_L}\sum_{j=1}^{d_R}A_{ij,ij}\right)^2
\le
\frac{(\Delta\hat{A})_\text{mc}^2}{D+1}
,
\label{eqn:TypicalPhi}
\end{align}
where $A_{ij,i'j'}=(\bra{E_i^{(L)}}\otimes\bra{E_j^{(R)}})\hat{A}(\ket{E_{i'}^{(L)}}\otimes\ket{E_{j'}^{(R)}})$ are the matrix elements of $\hat{A}$.
Replacing $\hat{A}$ by $\hat{\Omega}^\dag\hat{A}\hat{\Omega}$, these yield (\ref{eqn:AveMC}) and (\ref{secondorder1}). Then, the typicality of $\ket{\phi}_\infty$ and (\ref{eqn:PureNESSAve}) are proved for large $D$, if the microcanonical variance $(\Delta(\hat{\Omega}^\dag\hat{A}\hat{\Omega}))^2_\text{mc}$ does not scale badly in the thermodynamical limit.
One of the interesting and important quantities in the discussion of NESS is the current flowing between the reservoirs. The microcanonical variance of the current remains finite in the thermodynamical limit, since it is an intensive quantity. See the example in the next section.

\section{Example}
Let us look at an example.
We take \cite{Mahan1}
\begin{align}
\hat{H}_L&=\int d\omega\,\omega\hat{a}_\omega^\dag\hat{a}_\omega,\quad
\hat{H}_R=\int d\omega\,\omega\hat{b}_\omega^\dag\hat{b}_\omega,
\label{eqn:ModelHLR}
\\
\hat{V}&=\int d\omega\,d\omega'(g_{\omega,\omega'}\hat{a}_\omega^\dag\hat{b}_{\omega'}+g_{\omega,\omega'}^*\hat{b}_{\omega'}^\dag\hat{a}_{\omega})
\label{eqn:ModelV}
\end{align}
for the Hamiltonians in (\ref{eqn:Hamiltonian}), for the two weakly coupled reservoirs $L$ and $R$ [recall also Fig.\ \ref{fig:HirbertNESS}(a)].
Here, $\hat{a}_\omega$ and $\hat{b}_{\omega}$ are the canonical annihilation operators for the particles in reservoirs $L$ and $R$, respectively.
The following calculation is valid for both fermionic and bosonic cases.\footnote{We omit the spin degrees of freedom of the particles, since they are not essential to the following discussion.}
Without loss of generality, we assume that the density of states is common for $L$ and $R$ and the dispersion relations are linear. 
The particles in one reservoir are transferred to the other by the tunneling Hamiltonian $\hat{V}$.
The matrix elements of $\hat{V}$ are characterized by $g_{\omega,\omega'}$, whose finite and nonvanishing width in $\omega$ and $\omega'$ ensures the locality of the tunneling and regulates the momentum transfer in the tunneling process.
This locality is important for the current to be finite in the thermodynamical limit.
Reservoir $L$ is initially prepared at temperature $T_L$ and chemical potential $\mu_L$, while reservoir $R$ at $T_R$ and $\mu_R$, whose energies are given by $E_L$ and $E_R$, respectively.
In the long-time limit, the system approaches a NESS, in which a particle current flows between the two reservoirs steadily.

In the mathematical treatment, the energies $E_L$ and $E_R$ of the reservoirs are infinite in the thermodynamical limit.
It is implicitly assumed that we start with finite systems, and take the thermodynamical limit later, after the expectation value of an observable, e.g., of the current, is computed.
Note also that the long-time limit for $\hat{\Omega}$ in (\ref{eqn:Omega}) should be taken after the large-volume (continuum) limit.
See Footnote \ref{Note3} above (\ref{stationary1}) again.
Keeping these subtle points in our minds, we loosely write integrals and delta functions in the following formulas.
Mathematically rigorous treatment, e.g., based on the $C^*$-algebraic approach \cite{Tasaki2,Tasaki3,NESS-Ruelle,ref:NESS-AschbacherPillet-JSP,NESS-AschbacherJaksicPautratPillet,NESS-Tasaki}, should be discussed elsewhere.
Anyway, what is crucially important is the fact that the expectation value and the variance of the current operator in the microcanonical state are finite in the thermodynamical limit and in the long-time limit, which we already know from the standard ensemble approach to this problem \cite{Mahan1}.

Let us construct the typical pure NESS (\ref{stationary1}) for the present setting.
We sample a typical pure state $\ket{\phi}$ from $\mathcal{H}_{E_L,E_R}$, which is given in the form (\ref{eqn:Entanglement}), with $\ket{E_i^{(L(R))}}$ being the eigenstates of $\hat{H}_{L(R)}$, whose energies lie within the energy shell $[E_{L(R)},E_{L(R)}+\Delta E]$. They are actually given by the superpositions of Fock states.
We then apply the wave operator $\hat{\Omega}$ to construct a typical pure NESS $\ket{\phi}_\infty$.

We do it perturbatively with respect to the weak interaction $\hat{V}$.
To this end, it is convenient to recall the perturbative expansion for the wave operator $\hat{\Omega}$.
The wave operator $\hat{\Omega}$ in (\ref{eqn:Omega}) can be alternatively defined by $\hat{\Omega}=\lim_{\eta\downarrow 0}\eta\int_0^\infty dt\,e^{-\eta t} e^{-i\hat{H} t}e^{i \hat{H}_0 t}$ \cite{Thirring1}, which will coincide in the large-volume limit with the original $\hat{\Omega}$ defined in (\ref{eqn:Omega}), knowing that the long-time limit in (\ref{eqn:Omega}) exists.
The limit $\eta\downarrow0$ should be taken after the large-volume limit.
The wave operator $\hat{\Omega}$ defined in this way is cast into
\begin{equation}
\hat{\Omega}=1-\int dE\,\frac{1}{\hat{H}-E-i\eta}\hat{V}\delta(\hat{H}_0-E), \label{waveoperator}
\end{equation}
and the Dyson expansion of $1/(\hat{H}-E-i\eta)$ yields the perturbative expansion of $\hat{\Omega}$ as
\begin{align}
\hat{\Omega}
={}&1-\int dE\left(\frac{1}{\hat{H}_0-E-i\eta}\hat{V}\right.\nonumber \\
&\left.
{}-\frac{1}{\hat{H}_0-E-i\eta}\hat{V}\frac{1}{\hat{H}_0-E-i\eta}\hat{V}+\cdots
\right)\delta(\hat{H}_0-E). 
\label{perturbation1}
\end{align}
The higher-order terms describe multiparticle processes.

Using this formula for our model (\ref{eqn:ModelHLR})--(\ref{eqn:ModelV}), we compute $\ket{\phi}_\infty$ up to the first order in $\hat{V}$.
The typical pure state $|\phi\rangle$ in (\ref{eqn:Entanglement}) consists of the eigenvectors $|\Psi(E_L+E_R+\varepsilon_i)\rangle$ of $\hat{H}_0$ belonging to the eigenvalues $E_L+E_R+\varepsilon_i$, where $\varepsilon_i$ accounts for the energy relative to $E_L+E_R$. 
In the present case, the eigenstates are the superpositions of Fock states, and are quite degenerated in general.
We distinguish the degenerated eigenstates with different $i$'s.
Similarly, the vectors $\hat{a}_\omega^\dag\ket{\phi}$ ($\hat{b}_\omega^\dag\ket{\phi}$) and $\hat{a}_\omega\ket{\phi}$ ($\hat{b}_\omega\ket{\phi}$) are composed of the eigenvectors  belonging to the eigenvalues $E_L+E_R+\varepsilon_i+\omega$ and $E_L+E_R+\varepsilon_i-\omega$, respectively.  
Therefore, applying (\ref{perturbation1}) to $\ket{\phi}=\sum_{i}C_i|\Psi(E_L+E_R+\varepsilon_i)\rangle$, we get
\begin{align}
\ket{\phi}_\infty
=\ket{\phi}
&{}-\int d\omega\,d\omega'\,\biggl(
\frac{g_{\omega,\omega'}}{\omega-\omega'-i\eta}
\hat{a}_\omega^\dag\hat{b}_{\omega'}
\ket{\phi}
\nonumber\\
&\qquad\quad\ \ \,
{}+\frac{g_{\omega,\omega'}^*}{\omega'-\omega-i\eta}
\hat{b}_{\omega'}^\dag\hat{a}_{\omega}
\ket{\phi}
\biggr)
+\cdots.
\label{eqn:TypicalNESSModel}
\end{align}
Notice that $E_L+E_R+\varepsilon_i$, which are infinitely large in the thermodynamical limit, have disappeared.

Let us evaluate the current in this pure NESS\@.
The current is defined by the number of particles entering in reservoir $R$ per unit time \cite{Mahan1},
\begin{equation}
\hat{J}=-i[\hat{N}_R,\hat{H}]
=-i\int d\omega\,d\omega'(
g_{\omega,\omega'}^*\hat{b}_{\omega'}^\dag\hat{a}_{\omega}
-g_{\omega,\omega'}\hat{a}_\omega^\dag\hat{b}_{\omega'}
),
\label{eqn:Jdef}
\end{equation}
where $\hat{N}_R=\int d\omega\,\hat{b}_\omega^\dag\hat{b}_\omega$ is the number of particles in reservoir $R$. 
Its expectation value in the typical pure NESS $\ket{\phi}_\infty$ just constructed in (\ref{eqn:TypicalNESSModel}) is evaluated as 
\begin{widetext}
\begin{align}
\bras{\phi}{\infty}\hat{J}\ket{\phi}_\infty
\simeq{}&2\Im\int d\omega\,d\omega'
g_{\omega,\omega'}^*
\bra{\phi}\hat{b}_{\omega'}^\dag\hat{a}_{\omega}\ket{\phi}
\nonumber\\
&{}+2\Im\int d\omega_1\,d\omega_1'\int d\omega_2\,d\omega_2'\,
\frac{g_{\omega_1,\omega_1'}g_{\omega_2,\omega_2'}^*}{\omega_2'-\omega_2-i\eta}
\left(
\bra{\phi}\hat{a}_{\omega_1}^\dag\hat{b}_{\omega_1'}
\hat{b}_{\omega_2'}^\dag\hat{a}_{\omega_2}
\ket{\phi}
-
\bra{\phi}\hat{b}_{\omega_1'}^\dag\hat{a}_{\omega_1}
\hat{b}_{\omega_2'}^\dag\hat{a}_{\omega_2}\ket{\phi}
\right)
\nonumber\\
&{}-2\Im\int d\omega_1\,d\omega_1'\int d\omega_2\,d\omega_2'\,
\frac{g_{\omega_1,\omega_1'}^*g_{\omega_2,\omega_2'}}{\omega_2-\omega_2'-i\eta}
\left(
\bra{\phi}\hat{b}_{\omega_1'}^\dag\hat{a}_{\omega_1}
\hat{a}_{\omega_2}^\dag\hat{b}_{\omega_2'}
\ket{\phi}
-
\bra{\phi}\hat{a}_{\omega_1}^\dag\hat{b}_{\omega_1'}
\hat{a}_{\omega_2}^\dag\hat{b}_{\omega_2'}\ket{\phi}
\right)+\cdots.
\label{eqn:ExpJRaw}
\end{align}
\end{widetext}
\begin{floatequation}
\mbox{\textit{see eq.~\eqref{eqn:ExpJRaw}}}.
\addtocounter{equation}{-1}
\end{floatequation}
Here, we apply the typicality of $\ket{\phi}$ proved by (\ref{eqn:MicroPhi})--(\ref{eqn:TypicalPhi}) and the equivalence between the microcanonical and the grand canonical ensembles for large systems:
\begin{align}
&\quad\ %
\bra{\phi}\hat{b}_{\omega'}^\dag\hat{a}_{\omega}\ket{\phi}
\sim0,
\quad
\bra{\phi}\hat{b}_{\omega_1'}^\dag\hat{a}_{\omega_1}\hat{b}_{\omega_2'}^\dag\hat{a}_{\omega_2}
\ket{\phi}
\sim0, \label{matrixelement1}\\
&\bra{\phi}\hat{a}_{\omega_1}^\dag\hat{b}_{\omega_1'}\hat{b}_{\omega_2'}^\dag\hat{a}_{\omega_2}
\ket{\phi}
\nonumber\\
&\qquad 
\sim f_L(\omega_1)[1\mp f_R(\omega_1')]\delta(\omega_1-\omega_2)\delta(\omega_1'-\omega_2'),\label{matrixelement2}\\
&\bra{\phi}\hat{b}_{\omega_1'}^\dag\hat{a}_{\omega_1}\hat{a}_{\omega_2}^\dag\hat{b}_{\omega_2'}
\ket{\phi}
\nonumber\\
&\qquad
\sim [1\mp f_L(\omega_1)]f_R(\omega_1')\delta(\omega_1-\omega_2)\delta(\omega_1'-\omega_2'), \label{matrixelement3}
\end{align}
with $f_{L(R)}(\omega)=1/(e^{\beta_{L(R)}(\omega-\mu_{L(R)})}\pm1)$ for fermions (upper signs) and bosons (lower signs).  Even though the bosonic canonical operators $a_\omega$ and $b_\omega$ are unbounded operators, the typicality holds, since the relevant microcanonical variances are finite. Recall (\ref{eqn:TypicalPhi}).\footnote{Note that the delta functions are written loosely, as mentioned before. They are actually Kronecker's deltas for a finite volume of the system. In the thermodynamical limit, they become delta functions, but they accompany $g_{\omega,\omega'}$ in the current (\ref{eqn:Jdef}). The microcanonical variances of the relevant operators integrated with $g_{\omega,\omega'}$ are finite in the thermodynamical limit, and the typicality holds.
Equations (\ref{matrixelement1})--(\ref{matrixelement3}) are written keeping the presence of $g_{\omega,\omega'}$ in minds.}
Then, the expectation value of the current (\ref{eqn:ExpJRaw}) is reduced to 
\begin{equation}
\bras{\phi}{\infty}\hat{J}\ket{\phi}_\infty
\sim2\pi\int d\omega\,|g_{\omega,\omega}|^2
[
f_L(\omega)
-
f_R(\omega)
]+\cdots,
\end{equation}
where we have taken the thermodynamical limit and used $1/(x-i\eta)=\pv(1/x)+\pi i\,\delta(x)$.
This reproduces the Landauer-type formula derived by the ensemble approach \cite{Mahan1}.
The current flows with the difference between the Fermi/Bose distributions $f_L(\omega)$ and $f_R(\omega)$ of $L$ and $R$.

\section{Summary}
We have successfully constructed a class of typical pure states $\{|\phi\rangle_\infty\}$ which describe a NESS\@.
We just sample initial pure states randomly from $\mathcal{H}_{E_L,E_R}=\mathcal{H}_{E_L}\otimes\mathcal{H}_{E_R}$ to represent $\hat{\rho}_L\otimes\hat{\rho}_R$, and then ``scatter'' them to get a set of pure states, each of which describes the NESS, since the statistical error vanishes in the thermodynamical limit.      
In this way, the typicality of equilibrium states naturally amounts to the typicality of NESSs.

The thermodynamical structures of NESSs have been explored, e.g., by informational \cite{Antal1,Jaynes1} and dynamical approaches \cite{Tasaki2,Gaspard1}. 
The informational statistical mechanics suggests that NESSs can be characterized by the principle of maximization of an entropy under a constrained current \cite{Jaynes1}.
However, its applicability is restricted due to the omission of higher-order fluctuations. 
By using the Gibbsian state as the initial state, any higher-order fluctuations of current are fully taken into account \cite{Tasaki2,Gaspard1,Saito1,Nakamura,Esposito1}.
The typicality of pure NESSs allows us to relax the assumption of the initial Gibbsian state.
The equivalence (\ref{eqn:PureNESSAve}) holds also for the variance and the higher-order fluctuations of the current.
Note however that, if we look at all the moments of the current \textit{at the same time}, we acquire enough knowledge about the state to discriminate it from other pure NESSs and we lose the typicality: we should not look at too many moments. See \cite{Sugita1b,Sugita1,Reimann1}. It remains an important issue to be clarified better how many observables we can look at without losing the typicality.

Another interesting subject is the typicality under symmetry breaking. If we think of a generic many-body system, it might exhibit condensation at low temperatures. When a symmetry of a system is spontaneously broken, the naive random sampling from energy shells would not work, since in that way we might typically pick a superposition of states belonging to different phase sectors, which is not observed in real experiments. The typicality approach to statistical mechanics in the presence of symmetry breaking is an interesting and important future subject. An aspect relevant to this issue is discussed in \cite{ref:Typical-BEC-JC}.

Finally, the typicality of pure states would help reduce computational resources in evaluating thermodynamical quantities of large systems. Such an advantage is explored in \cite{Sugiura1,Sugiura2} for equilibrium systems. We expect the same to hold also for the typical pure NESSs, and the practical utility of typicality deserves a detailed study.

\acknowledgments
This work is supported 
by Grants-in-Aid for Young Scientists (B) (No.\ 26800206) and for Scientific Research (C) (No.\ 26400406) from JSPS, Japan, and by Waseda University Grants for Special Research Projects (2013A-982 and 2013B-147).


\end{document}